\documentclass[11pt]{article}
\usepackage{psfrag,epsfig}
\usepackage[latin1]{inputenc}

\usepackage{amsmath,amsfonts,latexsym, amssymb, amscd}

\newtheorem{satz}{Theorem}[section]

\newtheorem{lemma}[satz]{Lemma}
\newtheorem{koro}[satz]{Corollary}

\newtheorem{conclusion}[satz]{Conclusion}
\newtheorem{ob}[satz]{Observation}

\newcommand{\tit}{\textit}

\newcommand{\R}{\mathbb{R}}
\newcommand{\Z}{\mathbb{Z}}

\begin{document}
\thispagestyle{empty}
\begin{center}
\vspace*{1.0cm}

{\large{\bf Planck Fluctuations, Measurement Uncertainties and the\\
 Holographic Principle}}

\vskip 1.5cm

{\large {\bf Manfred Requardt}}\\email: requardt@theorie.physik.uni-goettingen.de 

\vskip 0.5 cm 

Institut f\"ur Theoretische Physik \\ 
Universit\"at G\"ottingen \\ 
Friedrich-Hund-Platz 1 \\ 
37077 G\"ottingen \quad Germany

\end{center}
\vspace{1 cm}
\vspace{1 cm}

\begin{abstract}Starting from a critical analysis of recently reported
  surprisingly large uncertainties in length and position measurements
  deduced within the framework of quantum gravity, we embark on an
  investigation both of the correlation structure of Planck scale
  fluctuations and the role the holographic hypothesis is possibly
  playing in this context. While we prove the logical independence of
  the fluctuation results and the holographic hypothesis (in contrast
  to some recent statements in that direction) we show that by
  combining these two topics one can draw quite strong and interesting
  conclusions about the details of the fluctuation structure and the
  microscopic dynamics on the Planck scale. We further argue that
  these findings point to a possibly new and generalized form of
  quantum statistical mechanics of strongly (anti)correlated systems
  of degrees of freedom in this fundamental regime.\\[0.3cm]
 \end{abstract} \newpage
\setcounter{page}{1}
\section{Introduction}
In recent years it has been argued that the at first glance quite
remote Planck scale might perhaps become observationally accessible by
devising certain ingeneous (thought) experiments. More specifically,
arguments were given that the quantum fluctuations of the space-time
metric or of distance measurements may come within the reach of
observability using already existing equipment like, for example, the
large and extremely sensitive interferometers, designed to detect
gravitational waves. This argument was in particular advanced by
Amelino-Camelia (see \cite{Amelino1} or \cite{Amelino2}) and supported
by various thought experiments and qualitative calculations given by
Jack Ng et al (see e.g. \cite{Ng1} or \cite{Ng2}). Interesting
arguments concerning the interface of general relativity and quantum
physics are also advanced in \cite{Ahluwalia}.

These arguments are provisional as a generally accepted theory of
quantum gravity does not yet exist and are of a character similar to
the quantum mechanical calculations before the advent of true quantum
mechanics in 1926. Nevertheless it is believed that they will hold in
a qualitative sense in any future theory of quantum gravity. In the
following we will also stick to this provisional reasoning.

The quantum gravity literature of the past decades abounds with such
heuristic arguments concerning the quantum behavior on the Planck
scale (see in particular the numerous remarks in \cite{Wheeler1} or
\cite{Wheeler2} or the paper by Padmanabhan, \cite{Padmanabhan}) with
the expected result that the relevant fluctuation effects are
essentially of Planck scale character. We note also the discussion in
e.g. \cite{Hossenfelder} where a possible minimal length is related to
large extra dimensions of space-time. One should however note that
the reasoning is not always complete as it is of course important to
take for example also the uncertainty in position of the (relevant
parts of the) measuring devices into account. We took some pains to
discuss this particular point in more detail in the following. In
contrast to these findings there are more recent arguments claiming
that some of these fluctuations (induced by quantum gravity) can
already be seen on a (in general) much larger scale, depending in a
somewhat surprising way also on the size of the quantities or objects
being measured. For example in \cite{Ng1} the uncertainty, $\delta l$,
of a length or distance measurement is claimed to go as
\begin{equation}\delta l\gtrsim l_p(l/l_p)^{1/3}        \end{equation}
with $l$ the length being measured and $l_p$ the Planck length.

The other interesting step consists of amalgamating this reasoning
with a version of the hypothetical \tit{holographic principle},
stating that on the Planck scale the number of degrees of freedom or
the information capacity of a spatial volume, $V$, go with the surface
area of $V$ and not! with the volume itself as in ordinary
(statistical) physics. A nice recent review is for example
\cite{Bousso} (we note that we plan to give a more complete list of
references elsewhere). It is claimed that the holographic principle,
stated in this particular form, supports the above findings
(\cite{Ng1}). In the following we want to scrutinize both lines of
reasoning and show that we come to different results. In this
connection we note that there are arguments that this particular form
of the holographic principle cannot hold in all possible situation. In
the following we mainly deal with weak gravitational fields and
relatively small but macroscopic subvolumes of practically infinite
space where these arguments do not apply. Anyway, we think that the
last word is not yet said on this subject matter as there exist
possible modifications of this relatively simple variant of the
principle also in more general situations. We will come back to this
question at the end of the paper.

We then proceed to show that both the fluctuation results and the
holographic hypothesis imply particular (anti)correlation constraints
of their own. By putting these two observations together we are able
to derive strong constraints on the correlations and dynamics of
degrees of freedom on the Planck scale. These findings seem to support
the view that in this fundamental regime a new or extended form of
statistical mechanics of strongly coupled or entangled degrees of
freedom and open systems may become necessary. We make some remarks in
this direction in the last section.

Concluding this brief r\'esum\'e we want to mention two recent papers
which discuss the interesting point of a possible change of particle
dispersion relations near the Planck scale and its consequences for
area laws and entropy bounds (\cite{Amelino3} and \cite{Camacho}). We
plan to come back to possible relations to our work elsewhere.
\section{A Discussion of Some Recent Results}
The first line of reasoning we mentioned above starts from an earlier
finding of Wigner et al (\cite{Wigner1}), dealing with quantum effects
concerning clocks and mirrors treated as test particles in a
gravitational field, $g_{ik}(x)$, and the setting-up of a coordinate
system on the space-time manifold. This line of thought is
supplemented by e.g. Ng et al by the wellknown argument concerning the
emergence of a black hole if too much mass is concentrated in a very
small region of space, the critical parameter being the
\tit{Schwarzschild radius}
\begin{equation}r_s=2GM/c^2        \end{equation}

Abbreviating the more detailed calculations in \cite{Wigner1}, one can
argue as follows. If one insists to measure distances in a
gravitational field by exchanging light signals between freely moving
small and sufficiently localized clocks and mirrors (tacitly assuming
that they do not disturb too much the given field $g_{ik}(x)$), the
following conclusion seems to be inevitable. We assume the clock (and
the mirror) initially to be localized with uncertainty $\delta
l$. Standard quantum mechanics leads to a momentum uncertainty
\begin{equation}\delta p\gtrsim \hbar/\delta l         \end{equation}
The average time, $\tau$, it takes for a light pulse to reach the mirror is
\begin{equation}\tau=l/c       \end{equation}
with $l$ the average distance between clock and mirror. In the time
interval $\tau$ the initial relative position uncertainty of clock and
mirror increases roughly as
\begin{equation}\delta l+\delta v\cdot\tau= \delta l+\hbar/m\delta
  l\cdot l/c=\delta l+\hbar l/mc\delta l     \end{equation}
with the minimum $\delta l$ being
\begin{equation}\delta l_{min}=(\hbar l/mc)^{1/2}      \end{equation}
(see \cite{Ng1}).

Quantum mechanics alone suggests to make the mass, $m$, of the clock (and
mirror) large in order to reduce the uncertainty in distance
measurement. Here now general relativity comes into the
play. Realizing for example the clock as a spherical cavity of
diameter $d$, surrounded by a mirrored wall of mass $m$, in which a
light signal bounces back and forth, the clock must tick off time at a
rate so that 
\begin{equation}d/c\lesssim \delta l/c      \end{equation}
in order that the uncertainty in distance measurement is not greater
than $\delta l$. On the other hand $d$ must be larger than the
Schwarzschild radius of the clock, $r_s$, so that signals can be
exchanged at all. This implies
\begin{equation}\delta l\gtrsim Gm/c^2      \end{equation}
Ng et al now combine these two estimates to get
\begin{equation}\delta l^3\gtrsim Gm/c^2\cdot \hbar l/mc   \end{equation}
or
\begin{equation}\delta l\gtrsim l_p(l/l_p)^{1/3}=(ll^2_p)^{1/3}
\end{equation}with $l_p=(\hbar G/c^3)^{1/2}$ the Planck
length. Correspondingly we get
\begin{equation}\delta\tau\gtrsim (\tau t_p^2)^{1/3}   \end{equation}
with $t_p=l_p/c$ the Planck time.

We briefly want to recapitulate how the \tit{true} spatial distance is
measured in a gravitational field in general relativity. This is
particularly clearly discussed in \cite{Landau}, see also \cite{Moeller}.
It comes out that for ``infinitesimal'' distances (which can in fact be
macroscopic in a sufficiently weak field for practical purposes) we
have
\begin{equation}dl^2=(g_{\alpha\beta}-g_{0\alpha}g_{0\beta}/g_{00})dx^{\alpha}dx^{\beta}=\gamma_{\alpha\beta}dx^{\alpha}dx^{\beta}>0      \end{equation}
with greek indices running from 1 to 3 and $\gamma_{\alpha\beta}$
being the spatial metric (the sign convention being $-+++$). We note
that in a physically realisable reference system we have $g_{00}<0$
(and corresponding constraints for the $\gamma_{\alpha\beta}$). Note
that in general gravitational fields the notion of true distance
(measured for example with little measuring rods or light signals) has
only an absolute meaning in the small. This is only different in
particular cases like a static field ($g_{ik}$ independent of the time
coordinate).

The crucial point in the analysis of Wigner et al was that clocks and
mirrors are treated as strongly localized freely moving test particles 
tracing out their individual world lines (or geodesics) in a given
gravitational field (similar discussions can e.g. be found in
\cite{Diosi} or \cite{Sasakura}). This is reasonable in a certain
context as e.g. in discussions of introducing appropriate coordinate
grids or material reference systems which do not distort too much the
given space-time. The situation however changes if questions of
principle are adressed in certain thought experiments concerning
fundamental limitations of e.g. length measurements in quantum gravity.
We argue in the following section that in that case some of the above
constraints can be avoided or at least relaxed so that the lower limit
provided by e.g. Ng et al can be considerably improved upon.
\section{An Alternative Thought Experiment}
We now describe a different set up which is not designed to create for
example a full coordinate grid or minimally disturb a given
gravitational field. We rather concentrate on the important question
of the existence of a priori limitations of measuring distances in the
very small with uncertainties much larger than $l_p$.

We note that a severe restriction in the approach of Wigner et al or
Ng et al derives from the assumption that clocks and mirrors are
freely moving particles of mass $m$. This resulted in the additional
($l$-dependent) position uncertainty $\hbar l/mc\delta l$. We
speculate what an experimenter in the laboratory would do. He would
fix clock and mirror on an optical bench, which we, for calculational
convenience, realize as a three-dimensional harmonic (macroscopic)
quantum crystal in a way described below. The unavoidable \tit{zero
  point motion} of the atoms of the solid is of the order
\begin{equation}\Delta q\sim (\hbar/M_0)^{1/2}            \end{equation}
with $M_0$ the mass of the atoms and $\Delta q$ their position
uncertainty (see e.g. \cite{Peierls}).

There exist various possibilities to implement the coupling between
clock, mirror and solid quantum mechanically. One possibility is to
confine both clock and mirror, as it is done with ordinary quantum
objects, in macroscopic (ionic or optical) traps which, on their part,
are attached to the solid. One may assume that these devices are
attached to a (macroscopic) part of the solid and not to a single
atom. This will yield an extra uncertainty of the order
$(\hbar/M)^{1/2}$ instead of $(\hbar/M_0)^{1/2}$ with $M$ the mass of
the respective part of the solid. 

 We approximate possible experimental
set-ups by assuming the clock (and mirror) to be bounded in the ground
state of a harmonic oscillator potential. This yields
\begin{equation}\Delta q^2=<x^2>=(m\omega/\hbar)^{1/2}\cdot\int
  x^2\exp(-m\omega x^2/\hbar)dx\sim (\hbar/m\omega)         \end{equation}
The momentum uncertainty
\begin{equation}\Delta p\gtrsim \hbar/\Delta q      \end{equation}
does now no longer matter as the particle cannot drift away
during the measurement process.\\[0.3cm]
Remark: The solid of course generates a gravitational field of its own
but we think, this does not represent a real problem in this context
as we are only interested in questions of principle. Assume for
example that the original metric was the Minkowski metric which is now
sligthly disturbed by the field of the solid.\vspace{0.3cm}

We now observe that by increasing $m$ and/or $\omega$, we still have
to obey the Schwarzschild constraint, but as the clock can not wander
away we get a bound like
\begin{multline}\delta l\gtrsim (Gm/c^2\cdot\hbar/m\omega)^{1/3}+(\hbar/M\omega)^{1/2}=(G\hbar/c^2\omega)^{1/3}+(\hbar/M\omega)^{1/2}\\=(l^2_pc/\omega)^{1/3}+(\hbar/M\omega)^{1/2}          \end{multline}
which does no longer contain an explicit $l$-dependence. $M$ is
possibly limited by practical or experimental constraints. But as the
respective region of the lattice need not be an infinitesimal one the
Scharzschild-constraint is at least not openly manifest.

At this place it is perhaps helpful to add a remark concerning another
deep question of principle which tacitly underlies all the discussions
of the kind presented above and similar ones but which, on the other
hand, is seldomly openly addressed. In quantum mechanics proper it
turned out that \tit{uncertainty in measurement} (Heisenberg) is
essentially the same as \tit{uncertainty in definability} (Bohr) which
mirrors sort of a preexisting harmony and is by no means a trivial
property from an epistomological point of view. As to this important
point cf. the discussion in \cite{Jammer} about the seemingly
different viepoints of Heisenberg and Bohr.

To put it briefly and relating it to our present problem concerning
the much more remote Planck scale: in our view it is not always
entirely obvious that every seeming limitation concerning the
measurement of a certain quantity like e.g. a distance by using a
particular measuring device really corresponds to a truely fundamental
limitation of definability of the quantity under discussion, that is,
as having its roots in for example irreducible primordial fluctuations
of space-time as such. To really decide this may be a touchy business
given the great recent advances in measurement techniques.
\section{Anticorrelated Space-Time Fluctuations}
In \cite{Ng1} it is argued that the $l$-dependent fluctuation formula
for length measurements, derived there, is further corroborated by an
application of the so-called \tit{holographic principle}. We want to
show in this and the following section that the holographic principle
is not really a cause for the given fluctuation formula but has rather
a logical status which is independent of that result. 

We start with a simple thought experiment concerning the nature of
Planck fluctuations which we presented already quite some time ago
(\cite{Req0}, we however presume that many other researchers in the
field are aware of this phenomenon). We assume that the quantum vacuum
on Planck scale is a fluctuating system behaving similar to systems in
quantum statistical mechanics with the characteristic correlation
parameters
\begin{equation}l_p=(\hbar
  G/c^3)^{1/2}\;,\;t_p=l_p/c\;,\;E_p=\hbar\nu_p=\hbar t_p^{-1}
\end{equation} We make, to begin with, the simplest possible but naive
assumption, assuming that in each Planck cell of volume $l_p^3$ we
have essentially independent energy fluctuations of size $E_p$ which
implies that the characteristic correlation length is assumed to be
$l_p$.

Picking now a macroscopic spatial volume (compared to the Planck
scale!), $V$, we have $N=V/l_p^3\gg 1$ of such cells labelled by $1\le
i\le N$. Defining the stochastic variable $E_V:=\sum_1^N E_i$, the
\tit{central limit theorem} tells us that the expected fluctuation of
$E_V$ is
\begin{equation}\Delta E_V:=<E_V\cdot E_V>^{1/2}\sim E_p\cdot N^{1/2}        \end{equation} 
where we assumed $<E_i>=0$ (a point we comment upon later). $\Delta
E_V$ would still be very large as both $N^{1/2}$ and $E_p$ are large
because $N$ itself is typically gigantic for macroscopic $V$. The
question is now, why are these volume-dependent large fluctuations not
observed?
\\[0.3cm]
Remark: In ordinary statistical mechanics extensive quantities like
e.g. $E$ go with the volume, $V$, or $N$. In that case it is
frequently reasonable to neglect fluctuations, being of order
$N^{1/2}$, as the scale used in our measurement devices is typically
adjusted to the occurring values of the extensive variables. Regarding
the Planck scale, we may however take the average of the vacuum energy
to be zero, or put differently, we do not measure it with our local
devices, but the fluctuations are expected to be large locally and
should be detectable in principle (see also the remarks in
\cite{Padmanabhan}).  \vspace{0.3cm}

We note that a similar reasoning yields for the momentum fluctuations
\begin{equation}\Delta p_V\sim p_{pl}\cdot N^{1/2}     \end{equation}
\begin{conclusion}On the Planck scale the hypothetical individual
  fluctuations must be strongly negatively or anti-correlated so that
  the integrated fluctuations in the volume $V$ are almost zero.
\end{conclusion}
We infer that what is called for are effective microscopic screening
mechanisms!

The above picture is of course quite crude but the same result would
essentially follow under much weaker and more realistic assumptions as
they are, for example, frequently made in (quantum) statistical
mechanics (cf. \cite{Req1}). To be specific, let $q(x)\,,\,x\in\R^d$
be a certain (quantum) observable density like e.g. some charge,
current or particle density. We normalize, for calculational
convenience, its expectation value, $\langle q(x)\rangle$, to zero. We
assume the system to be translation invariant and the \tit{correlation
  function}
\begin{equation}F(x-y):=\langle q(x)\cdot q(y)\rangle     \end{equation}
to be integrable, i.e.
\begin{equation}F(s)\in L^1(\R^d)    \end{equation}

With $Q_V:=\int_V q(x)d^dx$ the integral over a certain volume, $V$,
we get
\begin{multline}0\le\, \langle Q_VQ_V\rangle=\int_V dx\int_V dy F(x-y)=\int_V dx\left
    (\int_{x-V}ds F(s)\right )\\
\le\int_V dx\,sup|(\ldots)|\le \int_V
dx\cdot\int_{\R^d}ds\,|F(s)|=V\cdot const    \end{multline} 
as $\int_{\R^d}ds\,|F(s)|$ is finite by assumption. We can infer
the following:
\begin{conclusion}With $\langle q(x)q(y)\rangle\in L^1(\R^d)$ and 
\begin{equation}\lim_{V\to\R^d}\int_V F(s)\neq 0    \end{equation}
 we get
\begin{equation}\langle Q_VQ_V\rangle^{1/2}\sim V^{1/2}     \end{equation}
as in the case of complete independence of random fluctuations.
\end{conclusion}
Remark: Note that $N\sim V$ and summation is replaced by
integration over $q(x)$. The same reasoning holds of
course for discrete degrees of freedom. \vspace{0.3cm}

This Gaussian type of fluctuation can only be avoided if the
correlation function $F(s)$ displays a peculiar fine tuned
(oscillatory) behavior, more precisely, it must hold that
\begin{equation}\lim_{V\to \R^d}\int ds\,F(s)=0     \end{equation}
\begin{lemma}The rate of the vanishing of the above integral is
  encoded in the rate of vanishing of the Fourier transform,
  $\hat{F}(k)$, near $k=0$.
\end{lemma}
(see \cite{Req1} and \cite{Req2} and further references given
there). We hence can conclude that the behavior near $k=0$ of
$\hat{F}(k)$ is relevant for the degree of fluctuation of $Q_V$.

To establish this relation rigorously and also for other reasons
($q(x)$ is frequently not an operator function of $x$ but more
singular, i.e. only an operator valued distribution) it is customary
in quantum statistical mechanics and field theory to replace the sharp
volume integration over $V$ by a smooth scaling function. One may take
for example
\begin{equation}f_R(x):=f(|x|/R)      \end{equation}
with $f\geq 0\,\,f=1$ for $|x|\leq 1$ and being of compact support.
Instead of $Q_V$ we use in the following
\begin{equation}Q_R:=\int q(x)f_R(x)d^dx      \end{equation}
Remark: We note, without going into any details, that one can give
numerical estimates of the difference in behavior of the two
quantities. A certain disadvantage of a sharp volume cut off is that
it introduces an artificial \tit{non-integrability} in Fourier space
as the F.Tr. of a discontinuous function cannot be $L^1$! One can show
that in our case it is only in $L^2$.\vspace{0.3cm}

We get after Fourier transformation:
\begin{equation}\langle Q_R\cdot Q_R\rangle=R^{2d}\cdot\int d^dk\,\hat{F}(k)|\hat{f}(Rk)|^2     \end{equation}
Making a variable transform we get 
\begin{equation}\langle Q_R\cdot Q_R\rangle= R^d\int
  d^dk\,\hat{F}(k/R)|\hat{f}(k)|^2\sim const\cdot R^d\cdot
  R^{-\alpha}\cdot \int k^{\alpha}\cdot|\hat{f}(k)|^2d^dk       \end{equation}
asymptotically for $R\to\infty$ if 
\begin{equation}\hat{F}(k)\sim |k|^{\alpha}\quad\text{near}\quad k=0      \end{equation}
and vice versa. To show this we simply express $\hat{F}(k)$ as
$k^{\alpha}\cdot G(k)$ with $G(k)$ finite and nonvanishing at $k=0$. 
\begin{ob}Small or almost vanishing fluctuations in macroscopic volumes
  (compared to the Planck scale) can be achieved by certain covariance
  properties of the quantities under discussion, that is, a certain
  degree of vanishing of the F.Tr. of $\langle q(x)q(y)\rangle$ at
  $k=0$. For example, if $q(x)$ is the zero (charge) component of a
  conserved 4-current, we have typically a behavior $\sim |k|^2$ near
  $k=0$ for space dimension $d=3$. If $q(x)$ is the $00$-component of
  a conserved 2-tensor current, like e.g. the energy-momentum tensor,
  we have in general a behavior $\sim |k|^4$.
\end{ob}
(See \cite{Reeh} and \cite{Req2}).  

From the preceding discussion we hence can conclude that
small-fluctuations in $x$-space can obviously be achieved by
fine-tuned anticorrelations in $\langle q(x)q(y)\rangle$, which,
nevertheless, can be of short range, i.e. integrable.
\begin{koro}Small fluctuations as such in $V$ imply fine-tuned
  anticorrelations but not necessarily correlations having a long
  range.
\end{koro}
This is remarkable as we will show in the following that the
\tit{holographic principle} is intimately connected with long-range
correlations of a peculiar type. This implies that it is not a
necessary prerequisite for establishing small Planck
fluctuations.\\[0.3cm]
Remark: Recently Brustein et al (see for example \cite{Brustein1} and
\cite{Brustein2}) used field theoretic fluctuation results similar to
our results derived in e.g. \cite{Req1} and \cite{Req2} to argue that
such area-like scaling of fluctuations (occurring however in only very
particular situations) may be related to the area laws of the
holographic principle. We have to refrain in this letter-size format
to go into more details but plan to discuss this subtle point elsewhere.\vspace{0.3cm}

We conclude this section with providing a, as we think, instructive
example taken from ordinary physics which shows how easily these
strongly anticorrelated fluctuations appear even in non-relativistic
physics. We take again the 3-dimensional harmonic crystal mentioned
already above. We assume it to be fixed macroscopically in a definite
position, so that in the language of statistical mechanics its state
represent a \tit{pure phase}, in other words we assume a (spontaneous)
breaking of translation invariance. We concentrate in the following
for convenience on the atoms lying on the x-axis, their equilibrium
positions being the coordinates $\{j\cdot a\}\,,\,j\in\Z\,,\,a$ the
lattice constant. The momentary position of the $j$-th particle is
$x_j$. We know from statistical mechanics (cf. e.g. \cite{Peierls})
that the fluctuations of the microscopic particle positions are
finite (in three space dimensions!), i.e.
\begin{equation}\delta x_j^2=\langle(x_j-j\cdot a)^2\rangle<\infty     \end{equation}
for all $j$. On the other hand we have
\begin{equation}x_j-j\cdot a=\sum_{k=1}^j (x_k-x_{k-1})-ja+x_0         \end{equation}
with $<x_0>=0$.

The stochastic variables $(x_k-x_{k-1})=:u_k$ with $<u_k>=a$ play here
the role of the individual length fluctuations, $\delta l$, in the
corresponding Planck scale model. The crystal condition we assume to
be implemented by $\delta x_j^2\leq a^2$ (or a slightly weaker
condition) which is independent of $j$ due to the assumed translation
invariance.  We note that there exist various possibilities to
formulate such a condition but this is of no relevance for our present
discussion.

We have
\begin{multline}\label{A}   \delta
  x_j^2=\langle\left(\sum_{k=1}^j(u_k-a)+x_0\right)^2\rangle=\\
  \langle\sum_{k=1}^j(u_k-a)^2\rangle+\langle\sum_{k\neq
    k'=1}^j(u_k-a)(u_{k'}-a)\rangle +2\langle
  x_0\cdot\sum_1^j(u_k-a)\rangle + \langle x_0^2\rangle  \end{multline}
With both $\delta x_k^2$ and $\delta u_k^2$ independent of $k$ and
roughly of the same order, i.e. being $\lesssim (2a)^2$ we
see that, while the lhs of the equation is of order $(2a)^2$, the first
sum on the rhs is of order $j\cdot (2a)^2$. The third term on the rhs
can be calculated as follows. 
\begin{multline}\langle x_0\cdot\sum_1^j(u_k-a)\rangle=\langle
  x_0\cdot(x_j-x_0-ja)\rangle=\langle x_0\cdot(x_j-x_0)\rangle= \\
  \langle(x_0-\langle x_0\rangle)\cdot(x_j-\langle
  x_j\rangle)\rangle-\langle x_0^2\rangle \end{multline} as
$\langle x_0\rangle=0$ by assumption. In a pure phase we have clustering of
correlation functions, hence the first term on the rhs of the last
equation goes to zero for $j$ large. The third term of equation
(\ref{A}) is therefore of order $\sim\,\langle x_0^2\rangle$ for $j$
large and is hence compensated by the fourth term.

We hence arrive at the important constraint equation
\begin{equation}\langle\sum_{k\neq
    k'=1}^j(u_k-a)(u_{k'}-a)\rangle\approx -
  \langle\sum_{k=1}^j(u_k-a)^2)\rangle\lesssim-j\cdot (2a)^2   \end{equation}
\begin{conclusion}As $\delta x_j^2$ is globally bounded, $\delta
  x_j^2\lesssim a^2$, we find that the contributions in
  $\langle\sum_{k\neq k'=1}^j(u_k-a)(u_{k'}-a)\rangle$ are to a large
  part strongly negatively correlated in order to compensate the
  linear positive increase in $j$ of the first term on the rhs of
  equation (\ref{A}) (remember that $<u_k>=a$).
\end{conclusion}

Tranferring these observations to our Planck scale model, we can
associate $x_j$ or $j\cdot a$ with the momentary and averaged
macroscopical length we are going to measure, $\delta x_j$ with its
fluctuation and the $u_k$ with the individual but strongly
anticorrelated length fluctuations of the respective pieces of roughly
Planck size.  We see that an organized anticorrelation over large
length scales is obviously not entirely unnatural. In the example we
just studied it is connected with the occurrence of spontaneous
symmetry breaking (of translation invariance), that is, a phase
transition.  We hence conclude that it may happen under certain
circumstances that the fluctuation of a macroscopic length is of the
same order as the fluctuations of its much smaller parts in contrast
to what may be infered from a naive application of the central limit
theorem.
\section{Implications of the Holographic Principle}
In the preceding section we dealt with the possibility of small or
vanishing Planck scale fluctuations in macroscopic or mesoscopic
regions and consequences thereof. Our analysis led to \tit{strong
  anticorrelation result} but, as we saw, not necessarily to a
\tit{long-range} anticorrelation (see, however, the last example of
the harmonic crystal where exactly this happens to be the case). We
now add another aspect in form of the \tit{holographic principle}. We
employ it in the simple form typically used in various recent thought
experiments. That is, for simplicity reasons, we only deal with
situations where a \tit{space-like holographic bound} is supposed to hold
(cf. \cite{Bousso}).

As in the literature, cited above, we assume a given volume, $V$, to
be divided into $N=V/l_p^3$ or $N\sim V$ cells. We make the asumption
that each cell can store a finite amount of information, in the
simplest case one bit, represented by an internal state labelled by
the numbers $\pm 1$.

If these microscopic states can be independently chosen or more
realistically, as in the preceding section, are only finitely or
short-ranged correlated, we get an \tit{information storage capacity}
\begin{equation}I\sim V       \end{equation}
as it prevails in ordinary physics. The holographic principle claims
that on the Planck scale we have instead in certain situations a
behavior
\begin{equation}I\sim V^{2/3}\;\text{(surface
    area)}\;\text{or}\;I\sim l^2       \end{equation}
if $l$ is the diameter of $V$.

In \cite{Ng1} it is argued that such a behavior would naturally lead
to the length-fluctuation result reported there, i.e.
\begin{equation} \delta l\gtrsim l_p(l/l_p)^{1/3}     \end{equation}
We want to show in the following that this is, in our view, not the
case and that a different scenario is more plausible. 

In a first step we show that the holographic principle alone does not!
imply strong negative correlations among the microscopic degrees of
freedom. This property is rather the consequence of our above
\tit{small-fluctuation result}. If, for example, our microscopic
degrees of freedom are positively but long-range correlated in $V$,
this would, on the one hand, diminish the information storage capacity
(all spins or most of them are typically almost aligned in a given
microscopic fluctuation pattern in $V$), but an averaging over such
states would produce \tit{large global fluctuations} proportional to
(some fractional power of) $V$. But such large fluctuations are not
observed as was argued in the preceding sections.
\begin{conclusion}The holographic principle as such entails long-range
  correlations among the microscopic degrees of freedom in $V$
  (positive or negative ones). The small-fluctuation result, on the
  other hand, entails strongly negative correlations (but not
  necessarily of long range!). Taken together both principles entail
  strongly negatively correlated long-ranged fluctuations!
\end{conclusion}   

In \cite{Ng1} or \cite{Ng2} the holographic principle comes into the
play by arguing that the number of degrees of freedom in the volume
$V$ is on the one hand $l^3/\delta l^3$ and on the other hand
$l^2/l_p^2$, with $\delta l$ the minimal uncertainty of a length
measurement of $l$ (cf. the discussion at the beginning of this
paper). Ng argues that $\delta l$ is at the same time the minimal
length which can be resolved in $V$, put differently, which can be
attributed a physical meaning. He then relates the two expressions to
each other as
\begin{equation}l^3/\delta l^3\lesssim  l^2/l_p^2   \end{equation}
and gets 
\begin{equation}\delta l\gtrsim (l_p^2\cdot l)^{1/3}   \end{equation}
We however provided arguments in the preceding sections that
these expressions are not! logically related. 

We showed above that we may have $\delta l\sim l_p$ and hence in
principle (neglecting interactions among the cells) an available
number of degrees of freedom, $N=l^3/l_p^3$, without running into a
logical contradiction. The reason is that in our view the information
in $V$ is not!  stored in single more or less uncorrelated bits of
size $l_p^3$ or possibly varying size $\delta l^3$, but rather in a
strongly correlated pattern, extending over the full volume $V$. That
is to say, a configuration on the surface of $V$ can be freely chosen
(so to speak), leaving us with a total of roughly $2^{l^2/l_p^2}$
different surface states, i.e.  $I=l^2/l_p^2$, each of which induces,
due to long-ranged anti-correlations a more or less unique
configuration within $V$, extending this respective surface
configuration. As to this point see also the remarks in section 7 of
\cite{Susskind}.

This point of view has various consequences and ramifications also for
\tit{black hole physics} which we only briefly mention (see the
beautiful Galiean trialogue, \cite{Jacobson}). Suffice it to remark
that, adopting this point of view, there is no real problem in
combining individual length fluctuations of roughly Planck size,
$l_p$, with an information storage capacity going only with the
surface area of $V$.

We further note that the unique dependence of a volume state on a
corresponding surface state is not entirely unusual outside ordinary
(statistical) physics. Consider, for example, the \tit{Dirichlet
  property} employed in elliptic boundary value problems. With $L$ an
elliptic partial differential operator and $g(x)$ a configuration on
the boundary, $\partial V$, of $V$, there exists under fairly weak
conditions a unique solution, $f(x)$, in $V$, i.e. $Lf=0$, extending
smoothly to the boundary function, $g(x)$. If we discretize this
problem we get roughly an example with $I\sim |\partial V|$, where it
is understood that we label the volume states by the uniquely
associated surface states.

What we have said may provide a clue as to the more hidden reasons why
the holographic principle does not hold in the regime of the physics
of more ordinary length and energy scales. In ordinary quantum
statistical mechanics, for example, or when studying the
\tit{asymptotic distribution of eigenvalues} of elliptic partial
differential operators in a finite volume under selfadjoint boundary
condition (cf.  e.g. \cite{Hilbert}), the number of states below a
certain energy threshold $E$ is proportional to the phase-space volume
(see below). In this scenario we typically work with a fixed
(Hamilton) operator, $L$ (i.e. including fixed boundary conditions,
for example $f=0$ on $\partial V$) and study the eigenvalue problem
\begin{equation}Lf_i=\lambda_i f_i\;,\;\lambda_i\leq E   \end{equation}
with all $f_i$ being in principle physically admissible states.

We note that in statistical mechanics or theories of many degrees of
freedom in general we have to deal with Hamiltonians, describing the
interaction of many constituents (their number being typically
proportional to the ordinary geometric volume). In this case $V$ does
\tit{not} denote the ordinary geometric volume the system is occupying
but the generalized \tit{phase-space volume} which is a region having
a huge dimension and is rather of the order $e^V$ if $V$ denotes the
ordinary geometric volume. The information content is then of the
order $\log e^V=V$ (cf. e.g.  \cite{entanglement}).

Statistical mechanics tells us that one can in many cases regard a
subsystem, enclosed in the subvolume $V$, as being to some extent
independent of the ambient system which may be treated as a heat bath,
by neglecting the usually short ranged boundary effects or, rather,
incorporating them in a statistical manner. More specifically, one
makes for example a \tit{random phase approximation} in order to
arrive at a \tit{canonical partition function} with respect to the
subsystem contained in $V$ (see e.g. \cite{Tolman}). In such a
scenario, i.e. macroscopically excited states in the interior, the
entropy is typically proportional to the geometric volume. We note
that a similar behavior prevails if we study the
\tit{entanglement-entropy} of macroscopically excited eigenstates of a
Hamiltonian (describing the interaction of many degrees of freedom)
restricted to a subvolume (cf. \cite{entanglement}). That is, we can
make the following observation.

\begin{ob}In ordinary quantum statistical mechanics we study a
  subsystem contained in a subvolume, $V$, by neglecting to a certain
  extent the different microscopic boundary states the system can
  occupy, or, more specifically, by taking them only into account in a
  statistical way in, say, the canonical partition function, while we
  treat the bulk Hamiltonian as a fixed operator (i.e. fixed boundary
  conditions). We hence regard the possible correlations between the
  interior of $V$ and the heat bath outside as both sufficiently weak
  and irregular. We assume however that the internal degrees of
  freedom, that is, the eigenstates of $H$, can in principle all be
  excited  as their energy differences are assumed to be relatively
  small (this latter point being important in our view).  This yields
  a relation like $I\sim V$.
\end{ob}

Things however may change dramatically if the subsystem cannot really
be assumed to be separated from the ambient space due to very
\tit{long range} correlations between interior and exterior. We now
have to deal with a truely \tit{open subsystem}. In that case varying
boundary conditions have a strong effect on the interior of the system
and can no longer be emulated in a simple statistical overall manner,
quite to the contrary, in \tit{extreme} situations each possible boundary
state may induces a particular state in the interior.
\begin{ob}In this latter case we have to change our working
  hypothesis. Phrased in the language of statistical mechanics, we are
  no longer allowed to deal, on the one hand, with a fixed
  Hamiltonian, $H$, i.e. fixed boundary conditions together with the
  full sequence of its possible eigenstates, and incorporate the
  boundary fluctuations in the canonical partion function. Instead of
  that we rather may have to work with a fixed formal Hamilton
  operator for the interior of a given region supplemented with
  varying boundary conditions or, stated differently, a particular
  Hamiltonian for each boundary state. But it may now happen that we
  only see the respective ground states of this class of Hamiltonians
  being excited (or some few of the lowest lying of them) as the
  higher excited states may turn out to have a very high energy and
  are virtually not excited. In our Dirichlet example we thus may be
  allowed to take only the solutions
\begin{equation}Lf=0\cdot f\;,\;f=g\;\text{on}\;\partial V
\end{equation}
into account but now have to cover the full set of different possible
boundary conditions (or, rather, a countable set of typical ones).
\end{ob}    
Such a scenario would lead to an area-like behavior of entropy or
information content.\\[0.3cm]
Remark: We note that in supersymmetry breaking and other model
theories it is sometimes argued that the super-partners or higher
excited modes cannot be seen at ordinary energies due to their
supposed huge masses.\vspace{0.3cm}
\section{Conclusion}
We want to conclude this paper with a comment concerning seemingly
related findings in ordinary physics described by the notion of
entanglement-entropy (see \cite{entanglement} and the references given
there). We already mentioned that macroscopically excited eigenstates
of Hamiltonians describing the interaction of many degrees of freedom
lead in the generic case to partial states on subvolumes having an
entropy which is proportional to the volume while groundstates, away
from criticallity, that is in the regime of short-range correlations,
in the general case have an entanglement-entropy which goes with the
area of the boundary of the subvolume (note that there exist examples,
i.e. groundstates of product form, where it happens to be zero).

The holographic principle, on the other hand, deals with the maximally
possible entropy or information storage capacity of a region. That
means, we have to include also the highly excited states in our
considerations. But our above remark shows that in ordinary physics
these lead usually to a volume-behavior of
entanglement-entropy. Therefore we reach the following conclusion.
\begin{conclusion}The area law of the holographic principle cannot be
  understood within the context of ordinary (statistical) physics but
  needs prerequisites as described in the preceding sections, thus
  leading to a kind of generalized statistical mechanics.
\end{conclusion}

{\small
Acknowledgement: We thank the unknown referee for directing our
attention to the papers \cite{Amelino3} and \cite{Camacho}.\\[1cm]

}

\end{document}